\documentclass[aps,prl,reprint,superscriptaddress,amssymb,amsmath,floatfix]{revtex4-1}
\usepackage{graphicx}
\usepackage{hyperref}
\usepackage{color}
\usepackage[normalem]{ulem}
\usepackage{nicefrac}

\AtBeginDocument{
 \newwrite\bibnotes
 \def\bibnotesext{Notes.bib}
 \immediate\openout\bibnotes=\jobname\bibnotesext
 \immediate\write\bibnotes{@CONTROL{REVTEX41Control}}
 \immediate\write\bibnotes{@CONTROL{
  apsrev41Control,author="08",editor="1",pages="1",title="0",year="0"}}
 \if@filesw
  \immediate\write\@auxout{\string\citation{apsrev41Control}}
 \fi
}

\definecolor{cwh_color}{cmyk}{0, 1, 1, 0}

\begin{document}
\pdfsuppresswarningpagegroup=1

\title{Heisenberg spins on an anisotropic triangular lattice: PdCrO$_2$ under uniaxial stress}

\author{Dan Sun}
\affiliation{Max Planck Institute for Chemical Physics of Solids, N\"{o}thnitzer Str. 40, 01187 Dresden,
Germany}
\affiliation{Stewart Blusson Quantum Matter Institute,  the University of British Columbia, Vancouver,  V6T 1Z4 Canada}
\author{Dmitry A. Sokolov}
\affiliation{Max Planck Institute for Chemical Physics of Solids, N\"{o}thnitzer Str. 40, 01187 Dresden,
Germany}
\author{Richard Waite}
\affiliation{H. H. Wills Physics Laboratory, University of Bristol, Bristol, BS8 1TL, United Kingdom}
\affiliation{ISIS Facility, Rutherford Appleton Laboratory, Chilton, Didcot, OX11 0QX, United Kingdom}
\author{Seunghyun Khim}
\affiliation{Max Planck Institute for Chemical Physics of Solids, N\"{o}thnitzer Str. 40, 01187 Dresden,
Germany}
\author{Pascal Manuel}
\affiliation{ISIS Facility, Rutherford Appleton Laboratory, Chilton, Didcot, OX11 0QX, United Kingdom}
\author{Fabio Orlandi}
\affiliation{ISIS Facility, Rutherford Appleton Laboratory, Chilton, Didcot, OX11 0QX, United Kingdom}
\author{Dmitry D. Khalyavin}
\affiliation{ISIS Facility, Rutherford Appleton Laboratory, Chilton, Didcot, OX11 0QX, United Kingdom}
\author{Andrew P. Mackenzie}
\affiliation{Max Planck Institute for Chemical Physics of Solids, N\"{o}thnitzer Str. 40, 01187 Dresden,
Germany}
\affiliation{Scottish Universities Physics Alliance (SUPA), School of Physics and Astronomy, University of St. Andrews,
St. Andrews KY16 9SS, United Kingdom}
\author{Clifford W. Hicks}
\affiliation{Max Planck Institute for Chemical Physics of Solids, N\"{o}thnitzer Str. 40, 01187 Dresden,
Germany}
\affiliation{School of Physics and Astronomy, University of Birmingham, Birmingham B15 2TT, United Kingdom}

\date{\today}

\begin{abstract}

When Heisenberg spins interact antiferromagnetically on a triangular lattice and nearest-neighbor interactions dominate, the ground state is 120$^\circ$
antiferromagnetism. In this work, we probe the response of this state to lifting the triangular symmetry, through investigation of the triangular
antiferromagnet PdCrO$_2$ under uniaxial stress by neutron diffraction and resistivity measurements. The periodicity of the magnetic order is found to change
rapidly with applied stress; the rate of change indicates that the magnetic anisotropy is roughly forty times the stress-induced bond length anisotropy. At low
stress, the incommensuration period becomes extremely long, on the order of 1000 lattice spacings; no locking of the magnetism to commensurate periodicity is
detected.  Separately, the magnetic structure is found to undergo a first-order transition at a compressive stress of $\sim$0.4~GPa, at which the interlayer
ordering switches from a double- to a single-$q$ structure.

\end{abstract}

\maketitle

Antiferromagnetic interactions are frustrated on a triangular lattice. When the spins are Heisenberg spins and nearest-neighbour interactions dominate,
balancing interactions across the three bond directions leads to $120^\circ$ order, in which the spin orientation rotates by $2\pi/3$ from site to
site~\cite{Messio11_PRB, Poienar10_PRB}. Lifting the triangular symmetry may cause the magnetism to become incommensurate~\cite{Villain59,
Yoshimori59}.  However, propagation vectors tend to lock to commensurate values, because for commensurate orders the spins can find a single optimum
configuration, while for incommensurate orders the state energy must be averaged over a $2\pi$ rotation of the spin orientation. A symmetry-breaking field may
therefore need to exceed a threshold strength for the magnetism to become incommensurate. Such ``lock-in'' to commensurate periodicities has been observed in
several non-triangular systems, such as, for example, CaFe$_4$As$_3$~\cite{Manuel10_PRB, Nambu11_PRL}, Ba(Fe$_{1-x}$Co$_x$)$_2$As$_2$~\cite{Pratt11_PRL}, and
lightly-doped Cr~\cite{Fawcett94_RMP}; in all of these cases, the transition between commensurate and incommensurate states is first-order.

Independently of this potential commensurate-to-incommensurate transition, the possibility of tuning a triangular lattice through uniaxial stress has been
considered before, and there are two possible directions, illustrated in Fig.~1(a)~\cite{Villain59, Yoshimori59, Merino99, Zheng99}. By applying stress along a
$\langle \bar{1}10\rangle$ direction, the atomic positions are pushed towards a square lattice, and the magnetic interactions may similarly evolve towards a
square lattice with weak diagonal bonds. Alternatively, by applying stress along a $\langle 100 \rangle$ direction, the atomic positions are pushed in the
direction of 1D chains.

Here, we study the triangular antiferromagnet PdCrO$_2$ under tunable uniaxial stress applied along the $[\bar{1}10]$ lattice direction. We will show that
the magnetic interactions, like the atomic positions, evolve towards those of a square lattice. In other words, the interaction energy increases as bond length
is reduced, which is intuitive. In this compound, insulating CrO$_2$ layers are separated by highly conducting Pd sheets. The Cr configuration is $3d^3$, and
due to strong Hund's coupling the three electron spins align to form robust $S = 3/2$ moments; the CrO$_2$ layers are Mott insulating~\cite{Mackenzie17_RPP,
Sunko20_ScienceAdvances, Lechermann18_PRM}.  Coupling between the Cr and Pd layers means that the magnetic structure in the Cr layers affects electronic
transport properties. A downturn in resistivity shows that short-range correlation between Cr spins appears starting at around room
temperature~\cite{Takatsu09_PRB, Takatsu10_JPhysics, Hicks15_PRB}. By $\sim 100$~K the correlation length becomes long enough to measurably influence the Hall
and Nernst effects~\cite{Daou15_PRB}, and to yield substantial magnon drag in the thermopower~\cite{Arsenijevic16_PRL}. By 40~K the in-plane correlation area
reaches $\sim$1000~sites, without interlayer correlation~\cite{Billington15_ScientificReports, Ghannadzadeh17_NatComms}. Then, at $T_N = 37.5$~K, the layers
lock together in a transition that appears to be first-order, and the order becomes long-range correlated in three
dimensions~\cite{Billington15_ScientificReports}.

\begin{figure*}[ptb]
\includegraphics[width=150mm]{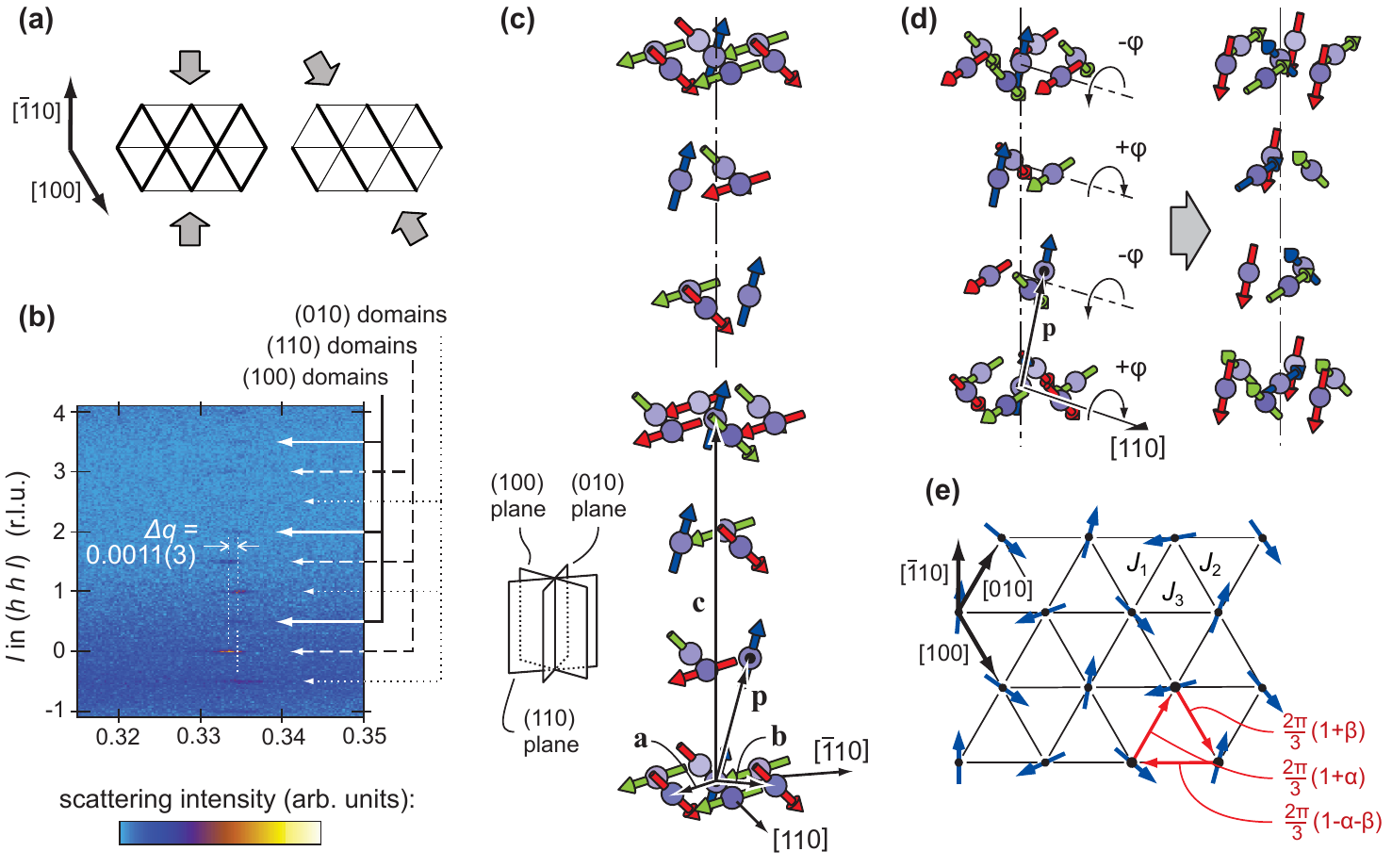}
\caption{\label{figBackground}
\textbf{(a)} Conceptual illustration of a triangular lattice under in-plane uniaxial stress. Heavy lines denote shorter bonds, and potentially stronger magnetic
interaction. Depending on the direction of the applied stress, the lattice can be pushed towards either a square lattice (left) or one-dimensional chains
(right). \textbf{(b)} Neutron scattering intensity in unstressed PdCrO$_2$, from Ref.~\cite{Sun19_PRB}. As indicated, there are three domain types, that yield
three sets of reflections. The offset $\Delta q$ is explained in the text. \textbf{(c)} A magnetic structure of PdCrO$_2$ consistent with the magnetic
scattering intensities reported in Ref.~\cite{Takatsu14_PRB}. The circles are the Cr sites, and the illustrated structure is from a $(110)$ domain.
$\mathbf{a}$, $\mathbf{b}$, and $\mathbf{c}$ are the unit cell vectors of the nonmagnetic lattice. $\mathbf{p}$ is the vector that connects interlayer
nearest-neighbor sites and lies within the spin plane; spins joined by $\mathbf{p}$ have the same colour. \textbf{(d)} Illustration of a collective spin
rotation that does not alter any magnetic scattering intensities. Here, $\varphi = 60^\circ$.  \textbf{(e)} An illustration of incommensurate order on a
near-triangular lattice. The red text indicates the spin rotation angle along the indicated bond directions, and $J_1$, $J_2$, and $J_3$ are the
nearest-neighbour magnetic interaction energies along the three bond directions.}
\end{figure*}

The magnetic order spontaneously lifts the three-fold rotational symmetry of the non-magnetic lattice~\cite{Hicks15_PRB, Sun19_PRB}. The delafossite structure
is rhombohedral: Cr sites in each layer are equidistant from three Cr sites in adjacent layers, so interlayer interactions balance and the layers cannot
couple~\cite{Komleva20_PRB}. The rotational symmetry breaking provides a preferred orientation that allows coupling; the related delafossite compound CuCrO$_2$
also shows this rotational symmetry breaking~\cite{Frontzek12_JPhysics, Soda09_PRB}.  The spin plane for PdCrO$_2$ and CuCrO$_2$ contains the $c$ axis, and the
azimuthal angle takes one of three symmetrically equivalent possible orientations~\cite{Takatsu14_PRB, Sun19_PRB}. 

We report results of neutron scattering and electrical resistivity measurements. The neutron diffraction data show that the rate of change of the magnetic
structure with applied stress is large: the magnetic anisotropy is roughly forty times the stress-induced bond length anisotropy. The neutron data also reveal a
stress-driven first-order transition in the interlayer ordering, which shows that the interlayer ordering probably reflects a delicate balance of interactions.
This transition is apparent in the resistivity data. The resistivity data also reveal structure in the transition at $T_\text{N}$, which encourage further
experimentation.

\section{Background}

We begin with a description of the magnetic structure of PdCrO$_2$ as it is currently understood. Shown in Fig.~1(b) is neutron scattering data from
nominally unstressed PdCrO$_2$, reported in Ref.~\cite{Sun19_PRB}. As indicated in the figure, there are three sets of reflections, originating from three
domain types, which we label by their spin plane. In Fig.~1(c), we illustrate a coplanar magnetic structure that gives a good fit to the scattering intensities
reported in Ref.~\cite{Takatsu14_PRB}. An anomalous Hall effect has been interpreted as evidence for a non-coplanar magnetic structure~\cite{Takatsu10_PRL}, and
an observation of nonreciprocal transport could indicate that inversion symmetry is broken~\cite{Akaike21_PRB}. However, the quality of the fit of the co-planar
structure in Fig.~1(c) is very good~\cite{Sun19_PRB}. The Hall effect might be influenced by magnetic breakdown~\cite{Ok13_PRL}. 

In the structure illustrated in Fig.~1(c), the spin plane is the $(110)$ plane. The $(100)$ and $(010)$ planes are symmeterically equivalent, resulting in
the three domain types. To understand the magnetic structure, it is useful to define a vector $\mathbf{p}$, that connects interlayer nearest-neighbour sites and
lies within the spin plane.  Between spins that are connected by $\mathbf{p}$, the components of the spins that are parallel to $\mathbf{p}$ are aligned
ferromagnetically, and those perpendicular to $\mathbf{p}$ are aligned antiferromagnetically.  This combination of ferromagnetic and antiferromagnetic alignment
means that the helicity alternates from layer to layer. It means that neutron reflections from a single domain are spaced along the $l$ axis by $\Delta l =
\frac{3}{2}$~r.l.u.  (defined using a three-layer unit cell), rather than the $\Delta l = 3$ that would be obtained for a fully ferromagnetic or
antiferromagnetic interlayer ordering.  For $h = k = \frac{1}{3}$, $(110)$ domains give reflections at $l = \frac{3}{2}n$, $(100)$ domains at $\frac{1}{2} +
\frac{3}{2}n$, and $(010)$ domains at $1 + \frac{3}{2}n$, where $n$ is an integar~\cite{Sun19_PRB}. 

\begin{figure*}[ptb]
\includegraphics[width=150mm]{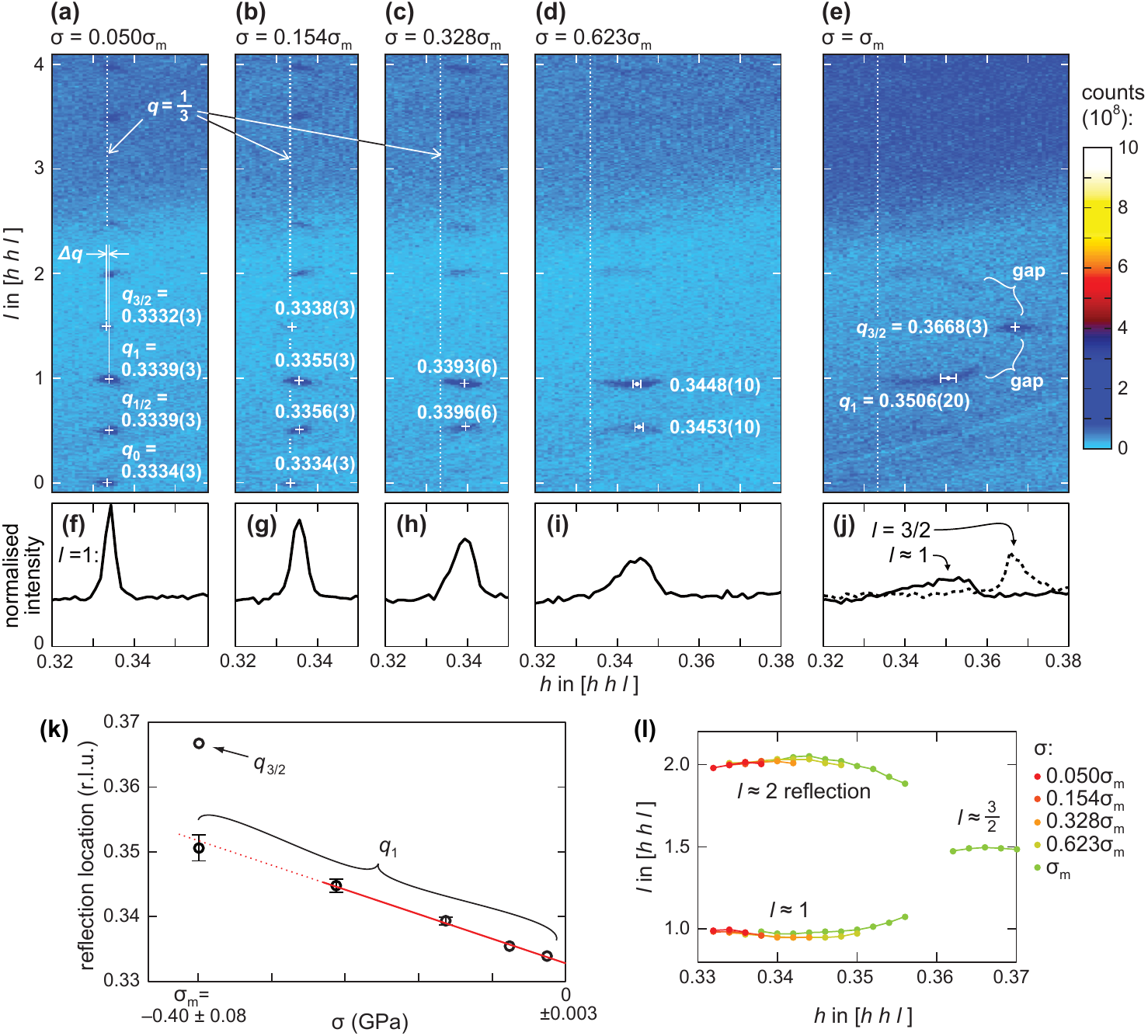}
\caption{\label{figHandLvsForce} Neutron scattering results, at 10~K. \textbf{(a--e)} Cuts through the $(h,h,l)$ plane at different uniaxial stresses $\sigma$
applied along the $[1 \bar{1}0]$ direction. The largest stress applied was $\sigma_m = -0.40 \pm 0.08$~GPa. The centres of selected reflections are indicated.
The peak shape varies with stress and is not straightforward to fit with a single model, so the centres and error bars are judged by eye.
These errors exclude any global error in the indexing.  All reflections are indexed according to the Bragg reflections at $\sigma = 0.050\sigma_m$~GPa.
\textbf{(f--j)} Cuts through the $l \approx 1$ reflections and, in panel (j), the $l = \frac{3}{2}$ reflection. In these plots, intensity is integrated over
$\pm 0.15$ in $l$ and $\pm 0.03$ in $[h, -h, l]$. \textbf{(k)} Location $q_1$ of the $l \approx 1$ reflection along the $h = k$ axis versus applied stress. The
line is a fit through the first four points; the dotted portion is an extrapolation.  \textbf{(l)} Paths traced by the $l \approx 1$, $l \approx 2$, and, at
$\sigma = \sigma_m$, $l \approx \frac{3}{2}$ reflections in the $(h,h,l)$ plane as stress is applied. The points mark the intensity maxima along cuts along the
$l$ axis.}
\end{figure*}

In Fig. 1(d) we illustrate a degree of freedom that is unconstrained by data so far: rotation of spins in alternating planes by angles $+\phi$ and $-\phi$
does not alter any neutron scattering intensities, for any value of $\phi$. Incommensuration can therefore emerge as a spatial gradient in $\phi$.  In Fig.
1(e) we illustrate a generic construction of incommensurate order on a triangular lattice. If the spin rotation angle upon translation from site to site along
the $[010]$ direction is $\frac{2\pi}{3}(1 + \alpha)$, and that along $[100]$ $\frac{2\pi}{3}(1 + \beta)$, then to linear order in $\alpha$ and $\beta$ the
reflection at $[h, k, l] = [\frac{1}{3}, \frac{1}{3}, l]$ moves to $[\frac{1}{3}(1+\alpha), \frac{1}{3}(1+\beta), l]$. In the illustration, the nearest-neighbor
interaction strengths are $J_1 > J_2 > J_3$, which yields $\alpha > 0$ and $\alpha + \beta > 0$.

\section{Results: neutron scattering}

A sample of cross section $1.37 \times 0.21$~mm$^2$ was mounted into piezoelectric-based uniaxial stress apparatus adapted for use in neutron scattering and
muon spin rotation measurements, described in Ref.~\cite{Ghosh20_RSI}. The central, exposed portion of the sample was 1.3~mm in length. The end
portions of the sample were embedded in epoxy for transmission of force to the sample. These portions were screened from the neutron beam with cadmium foil.  Growth
of single crystals of PdCrO$_2$ more than 1--2 mm in extent remains a challenge, and there were voids in the sample that will have introduced stress
inhomogeneity. 

Measurements were performed using the WISH diffractometer at the ISIS Pulsed Neutron and Muon Source~\cite{Chapon11_NeutronNews}. Compressive stress was
applied along the $[\bar{1}10]$ direction. Strain gauges were affixed to the sample holder to measure the applied stress, and the cell incorporates a mechanism
for finding the zero-stress reading. We estimate an error of $\pm 3$~MPa on this zero-stress reading. The largest compressive stress we applied was $\sigma_m =
-0.40 \pm 0.08$~GPa.  (Negative values denote compression.) Measurements were performed at 10~K, and temperature was kept constant, without thermal cycling,
during changes in the applied force. In analysis of the neutron data, we index the reflections at all strains to the Bragg peaks at the smallest stress
applied, $0.050\sigma_m$. In other words, there is no correction for lattice strain, which we show later to be a reasonable approximation.

Cuts through the $(h,h,l)$ plane (which is perpendicular to the stress axis) under different applied stresses $\sigma$ are shown in
Fig.~\ref{figHandLvsForce}(a--e). At $\sigma = 0.050\sigma_m$, reflections from all three domain types are present. The reflections from the $(100)$ and $(010)$
domains are shifted rightward relative to those from the $(110)$ domains by $\Delta q = 0.0006 \pm 0.0003$~r.l.u.  Such an offset can also be seen in the data
of nominally unstressed PdCrO$_2$ in Fig.~1(b), from Ref.~\cite{Sun19_PRB}, where $\Delta q = 0.0011 \pm 0.0003$~r.l.u. 

As stress is applied the $(110)$ reflections fade and disappear, which is in agreement with previous results~\cite{Sun19_PRB}: in-plane uniaxial compression
favors domains whose spin planes are more perpendicular to the stress axis. The $(100)$ and $(010)$ reflections shift further rightward. As $|\sigma|$ becomes large the
reflections also become streaked, which is most likely a consequence of strain inhomogeneity. Strain homogeneity should also make the nuclear Bragg peaks
streaked, but we find that $q$ shifts rapidly with applied strain, so that streaking is far more pronounced in the magnetic peaks. $q_1$, the location of the $l
= 1$ reflection, versus applied stress $\sigma$ is plotted in Fig.~2(k). Within resolution, the dependence is linear, and no locking to the commensurate value
$1/3$ is resolved.  Between $\sigma = 0$ and $-0.4$~GPa, $q_1$ shifts by 5\%. For a Young's modulus of $\sim$100~GPa, the longitudinal strain under $\sigma =
-0.4$~GPa would be around $-0.4$\%, and the Poisson's ratio expansion in the scattering plane perhaps half of this, or $+0.2$\%. In other words, the
stress-driven shift in $q_1$ exceeds plausible strain in the sample by an order of magnitude, and we are justified in not correcting the indexing for
lattice strain.

Because the spin planes of the $(100)$ and $(010)$ domains are neither aligned with nor perpendicular to the stress axis, there is no symmetry requirement for
the reflections from these domains to have $h = k$; referring to Fig.~1(e), we do not require $\alpha = \beta$.  In the Appendix cuts through the $(h, k, 1)$
plane are shown, which show that $h$ and $k$ of the reflections nevertheless remain approximately equal as stress is applied.

The fact that $\Delta q$ is non-zero at the smallest applied stress, and also in the nominally unstressed sample shown in Fig.~1(b), indicates that the magnetic
order is incommensurate even at very low applied stress, with an incommensuration period of around 1000 lattice spacings. To resolve such long-period
incommensuration, it is useful to be able to compare reflections from different domains in this way, because it means that we do not rely solely on the global
indexing of the reflections. A nonzero $\Delta q$ was visible in the data of Ref.~\cite{Sun19_PRB}, but was not commented upon: the observation that $\Delta q$
grows under stress gives confidence that it is a real feature of the data.

The paths traced by the $l = 1$ and $l = 2$ reflections in the $(h, h, l)$ plane as stress is applied are shown in Fig.~2(l). It is apparent that as stress is
applied the magnetism also becomes incommensurate along the $c$ axis. The reflection that was at $l=1$ shifts initially to an incommensurate position at
$l<1$, and then reverses direction to $l>1$ as $|\sigma|$ increases further. The $l = 2$ reflection follows an opposite stress dependence, shifting initially to
$l>2$, then $l<2$.

We do not detect any scattering intensity for $h$ between 0.357 and 0.367, where there is then a strong reflection at $l = \frac{3}{2}$. This gap is highlighted
in Fig. 2(e). We interpret these data as showing a first-order transition in the magnetic structure. In this interpretation, the reflections at $l \approx 1$
and $l \approx 2$ in the small-$|\sigma|$ structure shift continuously to larger $q$ as $|\sigma|$ is increased, eventually reaching $q = 0.357$. But magnetic
order is unstable for $0.357 < q < 0.367$. The low- and high-stress phases can coexist in equilibrium due to strain inhomogeneity. The cuts shown in
Fig.~2(j) support this interpretation: the $l \approx 1$ reflection has a tail extending to $h < 0.357$, and the $l = \frac{3}{2}$ has a tail extending to $h >
0.367$, but scattering intensity is cut off sharply between these two values.

The reflection $(0.367, 0.367, \frac{3}{2})$ is not accompanied by reflections at $l = 0$ and $l = 3$, showing that in the high-$|\sigma|$ phase the interlayer
correlation is strictly antiferromagnetic. Also, from the fact that the high-$|\sigma|$ reflection is at $l = \frac{3}{2}$, it appears that strong compression
along $[\bar{1}10]$ favors $(110)$ domains, whereas weak compression favored $(100)$ and $(010)$ domains.  The way that the reflections at $l \approx 1$ and
$l \approx 2$ curve inwards towards $l = \frac{3}{2}$ as $|\sigma|$ increases suggests a rotation of the spin planes of the $(100)$ and $(010)$ domains towards
the $(110)$ plane as stress is applied. However, the data here are not sufficient for accurate refinement of the magnetic structure under stress.

\section{Results: resistivity}

\begin{figure*}[ptb]
\includegraphics[width=145mm]{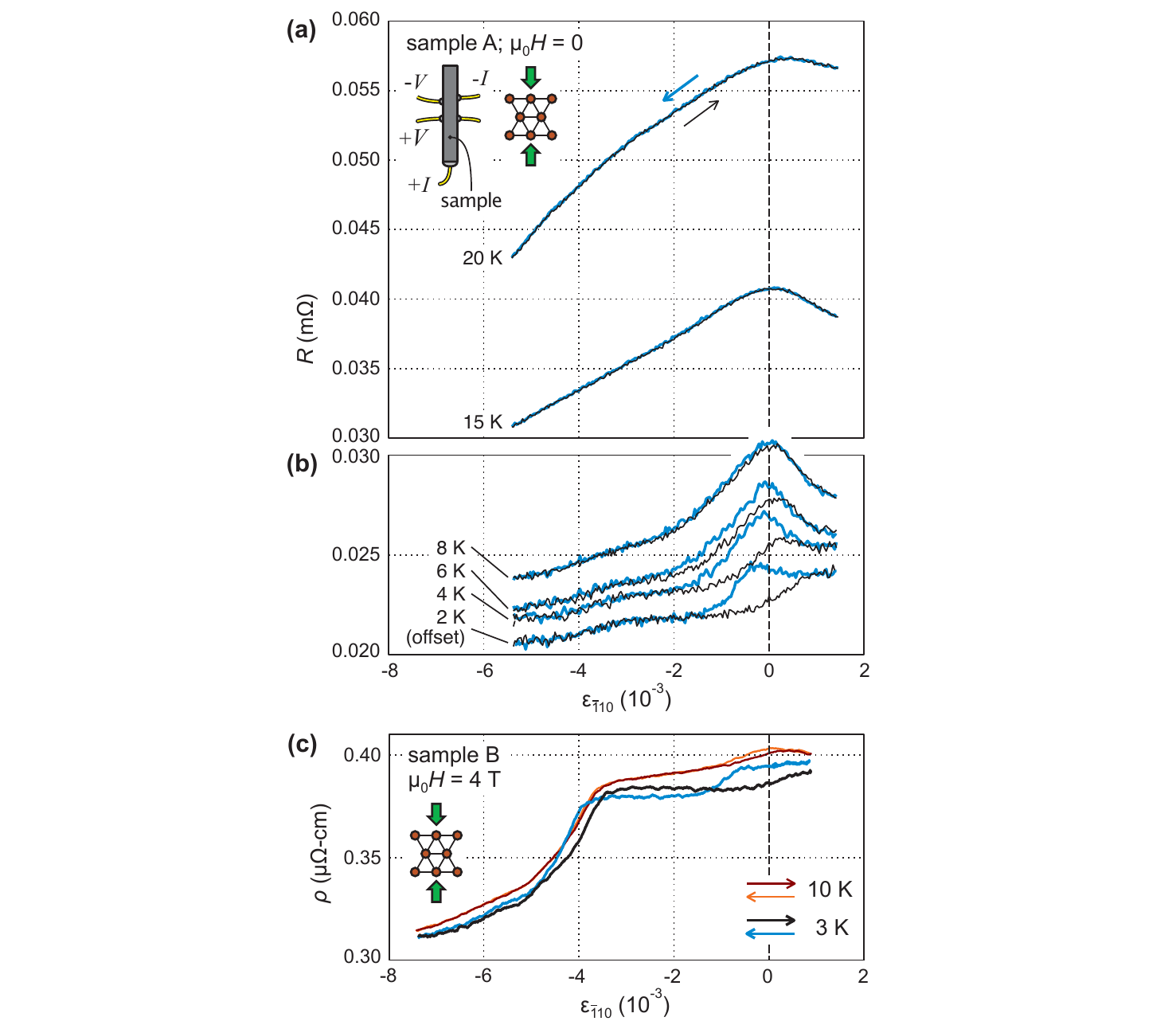}
\caption{\label{figLowTempTransport} \textbf{(a-b)} Resistance $R$ versus strain $\varepsilon_{\bar{1}10}$ for sample A at various temperatures. Note the
unconventional contact configuration, illustrated in the upper panel, due to a broken electrical contact. Data at 2~K are offset for clarity. \textbf{(c)}
Resistivity under an applied field along the $c$ axis $\mu_0 H = 4$~T for sample B, at 3 and 10~K.  The first-order nature of the transition at
$\varepsilon_{\bar{1}10} \approx -3.5\cdot10^{-3}$ is more apparent here than under zero field, in panel (a).}
\end{figure*}

We now show that this transition in interlayer ordering has an observable effect on in-plane resistivity. For resistivity measurements, a stress cell was
used that had a sensor of the displacement but not the force applied to the sample, so strain rather than stress is the measured control variable.  The applied
strain is taken as the applied displacement divided by the exposed sample length, and then multipled by a factor of $0.8$ as an estimate for the loss due to
elastic deformation of the epoxy.

Strain is applied along the length of the samples, and resistivity is measured along the same axis. Two samples were measured, labelled A and B.  Some data on
sample A were also shown in Ref.~\cite{Sun19_PRB}, where it was labeled Sample 1. In Fig.~\ref{figLowTempTransport}(a-b), we show the resistance versus
$\varepsilon_{\bar{1}10}$ for sample A at temperatures up to 20~K. (Resistance rather than resistivity is shown because a contact to the sample broke --- this
sample was small --- and so an unconventional contact configuration was used.) For temperatures below $\sim$ 8~K, the dominant feature in the data is a
first-order transition that was identified in Ref.~\cite{Sun19_PRB} as reorientation of domains as the applied stress switches between compressive and tensile,
and we therefore fix the neutral strain point $\varepsilon_{\bar{1}10}=0$ at the centre of this transition.

For temperatures below $\sim$8~K, a feature is also discernible at $\varepsilon_{\bar{1}10} \approx -3.5 \cdot 10^{-3}$. In the zero-field data from sample A it
is subtle, but under a field it becomes much more prominent: see the data from sample B in Fig.~\ref{figLowTempTransport}(c). It also becomes clear that it is a
first-order transition: there is definite hysteresis at 3~K, that mostly closes by 10~K. This is almost certainly the transition in interlayer ordering that is
seen in the neutron data. In the neutron data, the transition occurs at a stress of $\sim -0.4$~GPa, and if this is equated with the transition strain
$\varepsilon_{\bar{1}10} = -3.5 \cdot 10^{-3}$ seen in resistivity a Young's modulus of $\sim 100$~GPa is implied, which is a typical value for inorganic
solids.

\begin{figure*}[ptb]
\includegraphics[width=145mm]{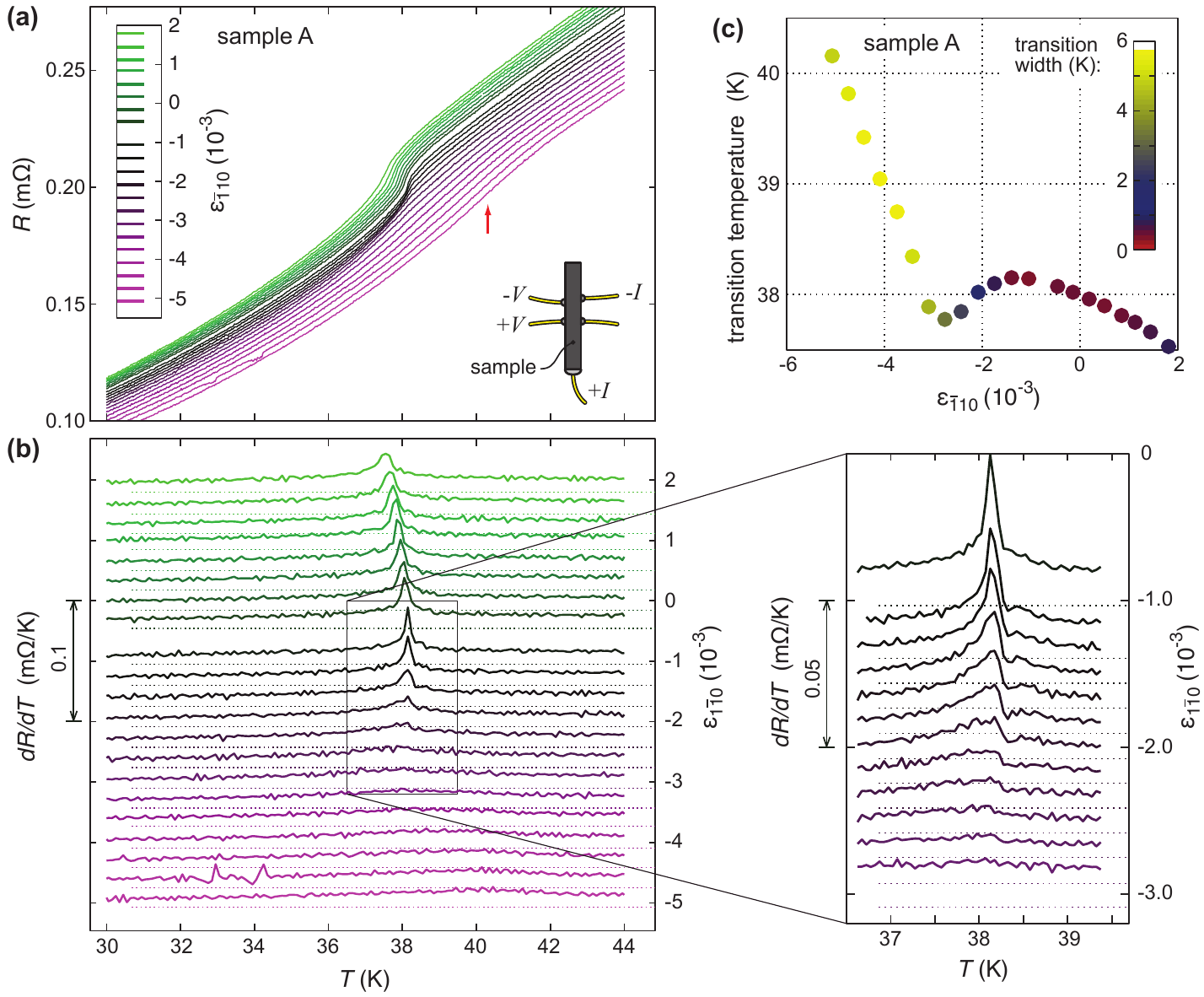}
\caption{\label{figHigherTempTransport} \textbf{(a)} $R(T)$ of sample A at various strains
$\varepsilon_{\bar{1}10}$. For small $|\varepsilon_{\bar{1}10}|$, $T_N$ is clearly discernible in the data. At larger $|\varepsilon_{\bar{1}10}|$,
the transitions broadens dramatically; at the maximum compression reached, $\varepsilon_{\bar{1}10} \approx -5 \cdot
10^{-3}$, the transition is visible as a broad maximum in the slope $dR/dT$, indicated by the red arrow.
\textbf{(b)} $dR/dT$ at various strains. The dotted lines show the applied strain, indicated
on the right-hand axis. \textbf{(c)} Transition temperature, identified as the temperature where $d^2R/dT^2 = 0$, versus
$\varepsilon_{\bar{1}10}$. The symbol color
indicates the width of the transition, taken as $[(1/R) \times |d^3R/dT^3|]^{-1/3}$ at the transition.}
\end{figure*}

Resistivity data also show that strain has a nontrivial effect on the N\'{e}el transition. $R(T)$ of sample A at $T \sim T_N$ and at various fixed strains
is shown in Fig.~4(a). At small applied strains, the N\'{e}el transition is apparent as a sharp drop in resistivity. Although no hysteresis was resolved, the
step-like nature of the transition strongly suggests that it is first-order. As the sample is compressed, there is initially little change in this feature,
but then it fades away. Under strong compression, a much broader feature is apparent in the data: a broad maximum in the slope $dR/dT$, indicated by the red
arrow in Fig.~\ref{figHigherTempTransport}(a) for the largest applied compression. Identifying this feature as the N\'{e}el transition, in Fig.~4(c) we show
$T_N$ versus strain, with the data points coloured by a measure of transition width. The crossover from a first-order to a broad transition is associated with a
change in the stress dependence of $T_N$: from a concave-downward, parabolic dependence to a steep linear dependence. Under tensile stress, the N\'{e}el
transition began to broaden, suggesting that it would evolve similarly under strong tensile strain as under strong compressive strain. However, the sample then
fractured as we attempted to tension it further.

\section{Discussion}

We begin our Discussion with consideration of the rate of change of the magnetic structure with strain. The fact that $q$ increases with compression
along $[\bar{1}10]$ means that the magnetic lattice, like the real-space lattice, is tuned towards a square net: referring to Fig.~1(e), $J_3$ shrinks relative
to $J_1$ and $J_2$. The largest stress we achieve, $\sigma_m = -0.4$~GPa, is modest. If the Young's modulus is $\sim$100~GPa, bonds along the stress axis would
be compressed by $\sim 0.4$\%, and bonds perpendicular to the stress axis stretched by $\sim$0.2\%, yielding a bond length anisotropy of $\sim$0.6\%. The
induced magnetic anisotropy appears to be much larger than this. At $\sigma = -0.4$~GPa, $q = 0.367$ for portions of the sample in the high-$|\sigma|$ state and
up to 0.357 for portions in the low-$|\sigma|$ state.  If interactions beyond nearest-neighbor are neglected, and $J_1$ and $J_2$ are taken to be equal, then $h
= 0.367$ is obtained for $J_3/J_1 = 0.746$, and $0.357$ for $J_3/J_1 = 0.803$~\cite{Villain59, Yoshimori59}--- the stress-induced anisotropy in nearest-neighbor
interactions is about forty times the bond length anisotropy. Higher-order interactions will alter these values, but the point is that magnetic anisotropy grows
very rapidly and so very substantial tuning of the magnetic structure is technically possible. For classical spins, spiral order is expected to undergo
transition to linear antiferromagnetism at $J_3 / J_1 = 0.5$~\cite{Yoshimori59}, and quantum corrections are predicted to shift this transition to higher
$J_3/J_1$~\cite{Merino99, Zheng99}. Extrapolation from the data here suggests that this transition should occur at $\sigma \sim -1$~GPa, which is an achievable
stress~\cite{Barber19_RSI}.

We did not reach such strong compression here, however, and the observed first-order transition, the change in interlayer ordering, was unexpected. It
suggests that interlayer ordering reflects a balance of interactions and so is somewhat delicate, an idea supported by the fact that other delafossite compounds
with CrO$_2$ layers have different interlayer orderings: in LiCrO$_2$, the helicity alternates from layer to layer~\cite{Kadowaki95_JPhysics}, as in unstressed
PdCrO$_2$, but in CuCrO$_2$ and AgCrO$_2$ the interlayer correlation is ferromagnetic~\cite{Kadowaki90_JPhysics, Frontzek12_JPhysics, Oohara94_JPSJ}, while, as
reported above, the interlayer ordering in PdCrO$_2$ under in-plane uniaxial stress is antiferromagnetic.

We discuss whether the magnetism of unstressed PdCrO$_2$ could be incommensurate.  Unstressed AgCrO$_2$ and CuCrO$_2$ are both incommensurate, with $q = 0.327$
for the former and $0.329$ for the latter~\cite{Oohara94_JPSJ, Poienar09_PRB, Frontzek12_JPhysics}. $q$ in unstressed PdCrO$_2$ appears to be much closer to
$\frac{1}{3}$ than in these materials. The spontaneous rotational symmetry breaking in unstressed PdCrO$_2$ will have a lattice distortion associated with
it, and it is possible that this distortion itself is sufficient to make the order incommensurate. Due to the possibility of residual thermal stresses on the
MPa level, both here and in~\cite{Sun19_PRB}, we cannot definitively resolve whether the magnetic order would be incommensurate under truly zero stress. Doing
so could be a subtle measurement challenge: with, for example, tetragonal to orthorhombic lattice distortions, a configuration of twin boundaries, parallel
lines, is possible that allows each domain to fully relax elastically, but with three domain types such an arrangement of domain walls is not possible.
Therefore, for measurement under truly zero stress, the sample would need to be polarised into a single domain.

Regardless of whether or not the magnetic order of unstressed PdCrO$_2$ is commensurate, the fact that the incommensuration period can reach 1000 lattice
spacings shows that the net spin-lattice coupling in PdCrO$_2$ is extraordinarily weak. It may be weak enough that classical dipole-dipole interactions
determine the orientation of the spin plane~\cite{Le18_PRB}.  The spins in PdCrO$_2$ appear to be almost perfect Heisenberg spins, which places PdCrO$_2$ in an
interesting regime for study of the thermodynamics of magnetic ordering. 

We conclude our Discussion with some notes on the effect of strain on the N\'{e}el transition, which provides hints of interesting dynamics in the formation
of magnetic order. As shown in Fig.~4(c), the first-order transition at $T_\text{N}$ extends out to $|\varepsilon_{\bar{1}10}| \sim 3 \cdot 10^{-3}$.  The
transition might be driven first-order by fluctuations: fluctuations are expected to drive rotational-symmetry-breaking transitions first-order when the
orientation is selected, as in PdCrO$_2$, from among three or more equivalent options~\cite{Hecker18_npj, Little20_NatMat}. The strain range over which the
transition is first-order does not correspond to fluctuations at $T < T_\text{N}$: the domain reorientation transition at $\varepsilon=0$ has been found to be
sharp for temperatures below, and right up to, $T_\text{N}$~\cite{Sun19_PRB}. It may therefore correspond to the strain required to polarise the fluctuations at
$T \gtrsim T_N$. In the mean field, the linear dependence of $T_\text{N}$ on uniaxial stress observed for $|\varepsilon_{\bar{1}10}| \gtrsim 3 \cdot 10^{-3}$ is
the expected form for transition into a uniaxial order. PdCrO$_2$ may therefore provide an interesting, and very well-controlled, test case for study of the
thermodynamic effect of fluctuations on ordering processes, which motivates further study of the transition at $T_\text{N}$.

\section{Acknowledgements} We thank C. Broholm, C. M. Lee, A. Little, T. Oka,  J. W. Orenstein, C. Timm, and M. Vojta for useful discussions. Experiments at the
ISIS Pulsed Neutron and Muon Source were supported by a beam time allocation from the Science and Technology Facilities Council under Expt. No.  RB1820290.
Financial support from the Deutsche Forschungsgemeinschaft through SFB 1143 (Project ID 247310070) and the Max Planck Society is gratefully acknowledged. RW
acknowledges funding and support from the Engineering and Physical Sciences Research Council (EPSRC) Centre for Doctoral Training in Condensed Matter Physics
(CDT-CMP), Grant No. EP/L015544/1.

\section{Appendix}

\subsection{Cuts in the $h, k, l=1$ plane}

\begin{figure*}[ptb]
\includegraphics[width=150mm]{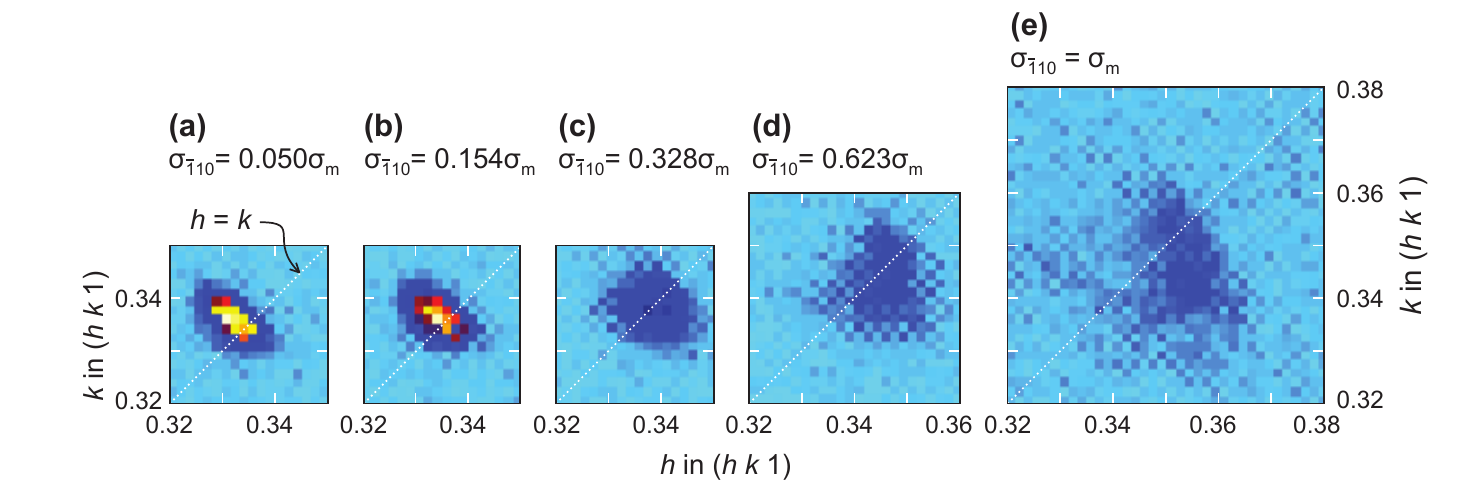}
\caption{\label{figHandKvsForce} Cuts through the $l \approx 1$ reflection in the $(h,k, l=1)$.}
\end{figure*}

In Fig.~\ref{figHandKvsForce}, we show cuts through the $l \approx 1$ reflection of Sample 1 in the $h,k$ plane. At the smallest applied stress, in panel (a),
the reflection lies close to the commensurate position $[\frac{1}{3}, \frac{1}{3}, l]$. As stress is applied, it moves outward to larger $h$ and $k$. It appears
not to follow the $h = k$ line precisely, though remains close enough that, within the resolution of these measurements, the cuts through the $(h, h, l)$ plane
in Fig. 2(a--e) give an accurate view of the evolution of the magnetic structure with stress.

\subsection{Error on the stress calibration for neutron measurements}

In the stress cell employed for neutron scattering, a mechanical contact can be opened between the piezoelectric actuators and the sample in order to bring the
force applied to the sample to nearly zero. However, the device incorporates a series of flexures, described in Ref.~\cite{Ghosh20_RSI}, and the spring constant
of these flexures, in combination with the differential thermal contraction between the sample and titanium, from which the sample holder is fabricated, can
still result in a residual stress on the sample. The total flexure spring constant is $\approx 0.3$~N/$\mu$m. A thermal contraction of 0.75\% between room and
cryogenic temperatures has been measured for PdCrO$_2$~\cite{Takatsu14_PRB}, but this is extremely large; CuCrO$_2$ contracts by 0.25\% between 300 and
10~K~\cite{Poienar09_PRB}. Ti contracts by 0.15\%, so we estimate a differential thermal contraction between PdCrO$_2$ and Ti of $\sim$0.1\%, and an effective
sample length over which this differential operates of $\sim$2~mm. The force on the sample would then be of order 2~$\mu$m $\times$ 0.3~N/$\mu$m $\sim$ 0.6~N,
for a stress of $\sim$3~MPa.

The strain gauges on the sample holder were calibrated after the neutron scattering measurement as described in Ref.~\cite{Ghosh20_RSI}. However, the response
from the gauges was roughly half the expected value, based on finite element analysis of the deformation of the holder under applied stress. We therefore assign
a relatively large error, $\pm 20$\%, to $\sigma_m$.

\nocite{apsrev41Control}
\bibliographystyle{apsrev4-1}
\bibliography{bibliography_settings,bibliography}

\end{document}